\documentstyle[12pt,epsfig]{article}
\begin{document}
\baselineskip=24pt
\def\rd{{\rm d}}
\newcommand{\nn}{\nonumber}
\newcommand{\ra}{\rightarrow}
\renewcommand{\baselinestretch}{1.5}
\begin{titlepage}
\vspace{-20ex}
\vspace{1cm}
\begin{flushright}
\vspace{-3.0ex} 
    {\sf ADP-01-53/T485} \\
\vspace{-2.0mm}
\vspace{5.0ex}
\end{flushright}

\centerline{\Large\sf Chiral Extrapolation of Lattice Data for the Hyperfine
Splittings} 
\centerline{\Large\sf of Heavy Mesons}
\vspace{6.4ex}
\centerline{\large\sf  	X.-H. Guo and A.W. Thomas}
\vspace{3.5ex}
\centerline{\sf Department of Physics and Mathematical Physics,}
\centerline{\sf and Special Research Center for the Subatomic Structure of
Matter,}
\centerline{\sf Adelaide University, SA 5005, Australia}
\centerline{\sf e-mail:  xhguo@physics.adelaide.edu.au,
athomas@physics.adelaide.edu.au,}
\vspace{6ex}
\begin{center}
\begin{minipage}{5in}
\centerline{\large\sf 	Abstract}
\vspace{1.5ex}
\small {Hyperfine splittings between the heavy vector ($D^*$, $B^*$)
and pseudoscalar ($D$, $B$) mesons have been calculated numerically
in lattice QCD, where the pion mass (which is related to the light quark mass) 
is much larger than its physical value. Naive linear chiral extrapolations
of the lattice data to the physical mass of the pion lead to hyperfine
splittings which are smaller than experimental data. In order to extrapolate 
these lattice data to the physical mass of the pion more reasonably, we apply 
the effective chiral perturbation theory for heavy mesons, which is invariant 
under chiral symmetry when the light quark masses go to zero and heavy quark 
symmetry when the heavy quark masses go to infinity. This leads to 
a phenomenological functional form with three parameters to 
extrapolate the lattice data. It is found that the extrapolated hyperfine 
splittings are even smaller than those obtained using linear extrapolation.
We conclude that the source of the discrepancy between lattice data for 
hyperfine splittings and experiment must lie in non-chiral physics.}

\end{minipage}
\end{center}

\vspace{0.5cm}

{\bf PACS Numbers}: 12.39.Fe, 12.39.Hg, 12.38.Gc, 12.40.Yx
\end{titlepage}

\vspace{0.2in}
{\large\bf I. Introduction}
\vspace{0.2in}

The past few years have seen much progress in lattice gauge theory, which is 
the only quantitative tool currently available to calculate nonperturbative  
phenomena in QCD from first principles. These phenomena include the
spectrum of light hadrons \cite{light}, the spectrum of heavy 
hadrons \cite{khan1, hein}, the decay constants of heavy mesons \cite{khan2},
nucleon structure functions \cite{gockeler}, and the Isgur-Wise function
for $B \ra D (D^*)$ transitions \cite{lacagnina}.

In Ref.\cite{hein}, the authors studied extensively the spectra of the $D$
and $B$ mesons using non-relativistic QCD (NRQCD) on the lattice in the
quenched approximation. For spin-independent splittings such as the
splittings between strange and non-strange mesons, good agreement with
experiments was obtained. However, it was found that the hyperfine 
splittings between $D (B)$ and $D^* (B^*)$ are much smaller than 
the experimental data. 
In fact, three lattice data for the mass of each of these mesons were
obtained in the region where the mass of the pion is larger than about
680MeV, which is much larger than the physical mass of the pion. With
naive linear extrapolations from the unphysical region to the physical
value of the pion mass, the extrapolated hyperfine  splitting between $D$ and 
$D^*$ is about 110MeV  for $\beta=5.7$ compared with the experimental
value 140MeV, whereas for $B$ and $B^*$ the extrapolated hyperfine  splitting
is about 29MeV compared with the experimental value 46MeV. Obviously these
large differences between the extrapolated and experimental values merit
careful investigation.

It is well known that QCD possesses a chiral $SU(3)_L \times SU(3)_R$  
symmetry in the limit where the masses of light quarks $u$, $d$, and $s$
go to zero. This symmetry is spontaneously broken into $SU(3)_V$, leading
to eight Goldstone bosons. The interactions of these pseudoscalar mesons  
are described by an effective chiral Lagrangian which is invariant under
$SU(3)_L \times SU(3)_R$. 
Another interesting quark mass limit in QCD is the heavy quark limit where
the masses of heavy quarks, $c$ and $b$, go to infinity. In this limit
there are heavy quark flavor symmetry and heavy quark spin symmetry
$SU(2)_f \times SU(2)_s$. Based on these symmetries which are not
manifest in the full theory of QCD, an effective theory for heavy quark
interactions which is called heavy quark effective theory (HQET) was 
established \cite{wise}. With the aid of HQET, the physical processes 
involving heavy quarks are greatly simplified. 
The interactions of heavy hadrons containing one heavy quark with the light
pseudoscalar mesons $\pi$, $K$ and $\eta$ should be constrained by
both chiral symmetry and heavy quark symmetry. The combination of these
two symmetries leads to an effective chiral Lagrangian for heavy hadrons
which is invariant under both  $SU(3)_L \times SU(3)_R$ and 
$SU(2)_f \times SU(2)_s$ transformations. There has been considerable
work in this direction in recent years \cite{hlcpt1, hlcpt2}.

In the past few years there has been a series of work dealing with 
extrapolations of lattice data for hadron properties, such as mass 
\cite{lein1}, magnetic moments \cite{lein2}, parton distribution functions
\cite{detmold}, and charge radii \cite{jones}, to the physical region. 
In these extrapolations the inclusion of pion loops yields leading and 
next-to-leading non-analytic behavior. This leads to rapid variation 
at small pion masses while lattice data are extrapolated to the physical
pion mass. However, when the pion mass is greater than some scale
$\Lambda$, which characterizes the physical size of the hadrons
which emit or absorb the pion, the hadron properties vary slowly and
smoothly. It is obvious that extrapolations which ensure the correct 
non-analytic behavior of QCD in this way should be more reliable
than a naive linear extrapolation.

With these considerations in mind, the aim of the present work is to 
extrapolate the lattice data in Ref.\cite{hein} to the physical region,
while building in the constraints of
chiral perturbation theory for heavy mesons, and to compare the extrapolated
hyperfine splittings with experiments.

The remainder of this paper is organized as follows. In Section II we 
give a brief review for chiral perturbation theory for heavy mesons.
In Section III we apply this theory to calculate self-energy 
contributions to heavy mesons from pion loops. Based on this, we  
propose a phenomenological, functional form
for extrapolating the lattice data for heavy meson masses to the  
physical region.  Then in Section IV we use this form to fit the
lattice data and give numerical results. Finally, Section V contains a 
summary and discussion.
 
\vspace{0.2in}
{\large\bf II. Chiral perturbation theory for heavy mesons}
\vspace{0.2in}

When the heavy quark mass $m_Q$ ($Q=b$, or $c$) is much larger than the
QCD scale $\Lambda_{\rm QCD}$, the light degrees of freedom in a heavy
hadron are blind to the flavor and spin orientation of the heavy quark $Q$.
Therefore, dynamics inside a heavy hadron remains unchanged under 
$SU(2)_f \times SU(2)_s$ transformations. In the opposite mass limit
$m_q \ra 0$, the QCD Lagrangian possesses an $SU(3)_L \times SU(3)_R$
chiral symmetry. The light pseudo-Goldstone bosons associated
with spontaneous breaking of chiral symmetry are incorporated in a
$3 \times 3$ matrix 
\begin{equation}
\Sigma = {\rm exp} \left( \frac{2iM}{f_\pi} \right),
\label{2a}
\end{equation}
where $f_\pi$ is the pion decay constant, $f_\pi=132$MeV, and  
\begin{eqnarray}
M=\left (
\begin{array}{ccc}
\frac{1}{\sqrt{2}}\pi^0+\frac{1}{\sqrt{6}}\eta & \pi^+ & K^+ \\
\pi^- &-\frac{1}{\sqrt{2}}\pi^0+\frac{1}{\sqrt{6}}\eta &K^0\\
K^- &\bar{K}^0 &-\sqrt{\frac{2}{3}}\eta\\
\end{array}
\right).     
\label{2b}
\end{eqnarray}
Under $SU(3)_L \times SU(3)_R$ transformations
\begin{equation}
\Sigma \ra L \Sigma R^+,
\label{2c}
\end{equation}
where $L \in SU(3)_L$ and $R \in SU(3)_R$. 

While discussing the interactions of Goldstone bosons with other matter
fields it is convenient to introduce 
\begin{equation}
\xi=\sqrt{\Sigma}.
\label{2d}
\end{equation}
Under $SU(3)_L \times SU(3)_R$ transformations
\begin{equation}
\xi \ra L \xi U^+=U\xi R^+,
\label{2e}
\end{equation}
where the unitary matrix $U$ is a complicated nonlinear function of $L$, $R$, 
and the Goldstone fields.

In order to discuss the interactions of Goldstone bosons with heavy
mesons, which consist of a heavy quark $Q$ and a light antiquark $\bar{q}^a$
($a=$1, 2, 3 for $u$, $d$, $s$ quarks respectively), 
a $4 \times 4$ Dirac matrix $H_a$ is introduced \cite{hlcpt1}
as follows:
\begin{equation}
H_a(v)=\frac{1+\rlap/v}{2}(P_a^{* \mu}\gamma_\mu-P_a \gamma_5),
\label{2f}
\end{equation}
where $P_a^{* \mu}$ and $P_a$ are field operators which destroy vector
and pseudoscalar heavy mesons with fixed four velocity $v$, respectively.
$P_a^{* \mu}$ satisfies the constraint
\begin{equation}
v_\mu P_a^{* \mu}=0.
\label{2g}
\end{equation}

Since $H_a$ is composed of a heavy quark and a light antiquark, under  
$SU(3)_L \times SU(3)_R$ 
\begin{equation}
H_a \ra H_b U^{+}_{ba},
\label{2h}
\end{equation}
and under heavy quark spin symmetry
\begin{equation}
H_a \ra S H_a,
\label{2i}
\end{equation}
where $S \in SU(2)_s$.

Defining
\begin{equation}
\bar{H}_a =\gamma_0 H_a^+ \gamma_0,
\label{2j}
\end{equation}
we have 
\begin{equation}
\bar{H}_a(v)=(P_{a \mu}^{* +}\gamma^\mu+P_a^+ \gamma_5)\frac{1+\rlap/v}{2},
\label{2k}
\end{equation}
which transforms as $\bar{H}_a \ra U_{ab} \bar{H}_b$ and $\bar{H}_a \ra 
\bar{H}_a S^{-1}$ under chiral symmetry and heavy quark symmetry, respectively.

It is convenient to introduce a vector field $V_{ab}^\mu$,
\begin{equation}
V_{ab}^\mu=\frac{1}{2}(\xi^+ \partial^\mu \xi +\xi \partial^\mu \xi^+)_{ab},
\label{2l}
\end{equation}
and an axial-vector field $A_{ab}^\mu$,
\begin{equation}
A_{ab}^\mu=\frac{1}{2}(\xi^+ \partial^\mu \xi -\xi \partial^\mu \xi^+)_{ab}.
\label{2m}
\end{equation}
Under $SU(3)_L \times SU(3)_R$, $V^\mu \ra U V^\mu U^+ +U \partial^\mu U^+$,
and $A^\mu \ra U A^\mu U^+$. Defining the covariant derivative
\begin{equation}
(D^\mu H)_a=\partial^\mu H_a -H_b V_{ba}^\mu,
\label{2n}
\end{equation}
we find that $D^\mu H \ra (D^\mu H) U^+$ under $SU(3)_L \times SU(3)_R$.
 
The Lagragian for the strong interactions of heavy mesons with Goldstone 
pseudoscalar
bosons should be invariant under both chiral symmetry and heavy quark
symmetry, since we are working in the limit where light quarks have
zero mass and heavy quarks have infinite mass. It should also be
invariant under Lorentz and parity transformations as required in general.
The most general form for the Lagragian satisfying these requirements is
\cite{hlcpt1}
\begin{equation}
{\cal L}= -{\rm Tr}[\bar{H}_a iv_\mu (D^\mu H)_a]+g {\rm Tr}(\bar{H}_a
H_b \gamma_\mu A^\mu_{ba} \gamma_5),
\label{2o}
\end{equation}
where $g$ is the coupling constant describing the interactions between
heavy mesons and Goldstone bosons. Obviously, $g$ is universal for
$D$, $B$, $D^*$, and $B^*$. Since $g$ contains information about the
interactions at the quark and gluon level, it cannot be fixed from chiral
perturbation theory for heavy mesons, but should be determined by 
experiments.  

From Eq.(\ref{2o}), the propagator for the pseudoscalar meson $D$ (or $B$) is 
$$\frac{i}{2v \cdot p}$$ 
where $p$ is the residual momentum of the
meson. For the vector meson $D^*$ (or $B^*$), the propagator is
$$\frac{-i(g_{\mu\nu}-v_\mu v_\nu)}{2v \cdot p}.$$ 
In the limit $m_Q \ra \infty$, there is no mass difference between 
pseudoscalar and vector mesons. 

Since in this work we will study hyperfine splittings,
we need to take $1/m_Q$ corrections into account. At order $1/m_Q$
in HQET, the term which is responsible for hyperfine splittings is 
the color-magnetic-moment operator, $\bar{h}_v \sigma_{\mu\nu}G^{\mu\nu} h_v$
(where $h_v$ is the heavy quark field operator in HQET and $G^{\mu\nu}$
is the gluon field strength tensor). This leads to the following correction 
term to ${\cal L}$ in Eq.(\ref{2o}):
\begin{equation}
\frac{\lambda_2}{m_Q} {\rm Tr} \bar{H}_a \sigma^{\mu\nu} H_a \sigma_{\mu\nu},
\label{2p}
\end{equation}
where $\lambda_2$ is a constant which also contains interaction information
at the quark and gluon level, and which is same for $D$, $B$, $D^*$, and $B^*$
at the tree level. (When QCD loop corrections are included, $\lambda_2$ 
depends on $m_Q$ logarithmically.) Finally, we note 
that the inclusion of the other term at
order $1/m_Q$ in HQET, $\frac{1}{m_Q}\bar{h}_v (iD)^2 h_v$, 
leads to a slight $m_Q$ dependence of the coupling constant $g$. 

Adding the term (\ref{2p}) to the Lagrangian for HQET, Eq.(\ref{2o}),
and using Eqs.(\ref{2f}) and (\ref{2k}), we can easily see that the mass
difference between vector and pseudoscalar heavy mesons is 
\begin{equation}
\Delta=-\frac{8\lambda_2}{m_Q},
\label{2q}
\end{equation}
and consequently, the propagators for heavy pseudoscalar and vector mesons
become 
$$\frac{i}{2(v \cdot p +\frac{3}{4}\Delta)}$$ 
and 
$$\frac{-i(g_{\mu\nu}-v_\mu v_\nu)}{2(v \cdot p -\frac{1}{4}\Delta)},$$
respectively.

In order to consider the interactions of heavy mesons with Goldstone bosons,
we substitute $\xi={\rm exp}(iM/f_\pi)$ into Eqs.(\ref{2l}, \ref{2m})
and obtain the following expressions for  $V^\mu$ and $A^\mu$:
\begin{equation}
V_\mu=\frac{1}{2f_\pi^2}[M, \partial_\mu M] + O(M^4),
\label{2r}
\end{equation}
\begin{equation}
A_\mu=-\frac{1}{f_\pi} \partial_\mu M +O(M^3).
\label{2s}
\end{equation}

Substituting Eq.(\ref{2r}) and  Eq.(\ref{2s}) into Eq.(\ref{2o}) we have the
following explicit form for the interactions of heavy mesons with Goldstone
bosons:
\begin{eqnarray}
{\rm Tr}[\bar{H}_a iv_\mu V^\mu_{ba} H_b]+g {\rm Tr}(\bar{H}_a
H_b \gamma_\mu A^\mu_{ba} \gamma_5)&=&\frac{i}{f_\pi^2}v^\mu
[M,\partial_\mu M]_{ba}(P_{a \nu}^{* +} P_{b}^{* \nu}- P_a^+ P_b) \nn\\
&& -\frac{2g}{f_\pi}(P_{a \mu}^{* +} P_b \partial^\mu M_{ba}
+P_a^+ P_{b \mu}^{*}\partial^\mu M_{ba} \nn\\
&&+i \epsilon^{\mu\nu\rho\sigma}
P_{a \rho}^{* +}P_{b \sigma}^{*}v_\nu \partial_\mu M_{ba}),
\label{2t}
\end{eqnarray}
where $O(M^3)$ terms are ignored.

Chiral symmetry can also be broken explicitly by nonzero light quark masses.
Under $SU(3)_L \times SU(3)_R$, light quark mass terms transform as
$(\bar{3}_L, 3_R) + (\bar{3}_R, 3_L)$. Then, to leading order in the
explicit chiral symmetry breaking from light quark masses, the following
terms are added to the Lagrangian in Eq.(\ref{2o}):
\begin{equation}
\lambda_1 {\rm Tr}\bar{H}_b H_a (\xi m_q \xi +\xi^+ m_q \xi^+)_{ab}
+\lambda_1^{\prime} {\rm Tr}\bar{H}_a H_a (\xi m_q \xi +\xi^+ m_q \xi^+)_{bb},
\label{2u}
\end{equation}
where $\lambda_1$ and $\lambda_1^{\prime}$ are parameters which are also
independent of the heavy quark mass in the limit $m_Q \ra \infty$.
 
\vspace{0.2in}
{\large\bf III. Formulas for the extrapolation of heavy meson masses}
\vspace{0.2in}

In this section we use the Lagrangian for the interactions of heavy
mesons with light Goldstone bosons to calculate pion loop corrections
to the masses of heavy vector and pseudoscalar mesons. This yields
the dependence of heavy meson masses on the pion mass. Then we 
propose a phenomenological functional form for extrapolating lattice
data for heavy meson masses to the physical pion mass.

From Eq.(\ref{2t}) we can see that there are five possible diagrams
for pion loop corrections to heavy meson masses. These diagrams are shown
in Fig.1 for $D$ and $B$ mesons, and in Fig.2 for $D^*$ and $B^*$ mesons.
Fig.1(a) and Fig.2(a) arise from the first term in Eq.(\ref{2t}). It can
be easily seen that these two diagrams do not contribute to the
masses of heavy mesons and we will not consider them from now on.

Fig.1(b) represents the pion loop correction to the heavy pseudoscalar meson
propagator ($P$ could be $D$ or $B$). 
In momentum space it can be expressed as
\begin{equation}
\frac{i}{2(v \cdot p +\frac{3}{4}\Delta)} (-2i\Sigma)
\frac{i}{2(v \cdot p +\frac{3}{4}\Delta)},
\label{3a}
\end{equation}
where $p$ is the residual momentum of the pseudoscalar meson and 
$-2i\Sigma$ is given by the following integral:
\begin{equation}
-2i\Sigma=-\frac{3 g^2}{f_\pi^2}\int \frac{{\rm d}^4 k}{(2\pi)^4}
\frac{k^2-(v \cdot k)^2}{[v \cdot (p-k)-\frac{1}{4}\Delta](k^2-m_\pi^2)},
\label{3b}
\end{equation}
where $k$ is the momentum of the pion in the loop, and $m_\pi$ is the
pion mass which is not necessarily its physical mass.
It can be seen that $\Sigma$ is a function of $v \cdot p$, $m_\pi^2$, and 
$\Delta$. After the correction from $\Sigma$ is added, 
the heavy meson propagator is proportional to
\begin{equation}
\frac{1}{v \cdot p -m_0 - \Sigma(v \cdot p)},
\label{3c}
\end{equation}
where $m_0$ is the mass term without $\Sigma$ correction (for the propagator
of the pseudoscalar heavy mesons, $m_0=-\frac{3}{4}\Delta$). 
The physical mass of the heavy meson, $m$, is defined by
\begin{equation}
[v \cdot p -m_0 - \Sigma(v \cdot p)]|_{v\cdot p =m}=0.
\label{3d}
\end{equation}
Therefore, to order $O(g^2)$ we have
\begin{equation}
m= m_0 + \Sigma(v \cdot p =m_0).
\label{3e}
\end{equation}

In order to calculate the integral in Eq.(\ref{3b}), we need to deal with
the following integral:
\begin{equation}
X^{\mu\nu} \equiv \int \frac{{\rm d}^4 k}{(2\pi)^4}
\frac{k^\mu k^\nu}{[v \cdot k - \delta](k^2-m_\pi^2)},
\label{3f}
\end{equation}
where $\delta=v \cdot p -\frac{1}{4}\Delta$ for Eq.(\ref{3b}).
On the grounds of Lorentz invariance, in general we have
\begin{equation}
X^{\mu\nu} = X_1 g^{\mu\nu}+ X_2 v^\mu v^\nu,
\label{3g}
\end{equation}
where $X_1$ and $X_2$ are Lorentz scalars, which are functions of 
$v \cdot p$, $m_\pi^2$, and $\Delta$. It can be easily seen from 
Eq.(\ref{3b}) that $\Sigma \sim (g^{\mu\nu}-v^\mu v^\nu)X_{\mu\nu}$. 
Consequently, only the $X_1$ term contributes, since $v^2=1$. 
Multiplying both sides of 
Eq.(\ref{3f}) with $(g^{\mu\nu}-v^\mu v^\nu)$ we have
\begin{equation}
X_1 =\frac{1}{3}\int \frac{{\rm d}^4 k}{(2\pi)^4} \frac{k^2-(v \cdot k)^2}
{v^0 (k_0-\frac{{\bf v}\cdot{\bf k}+\delta}{v^0}-i\epsilon)(k_0^2
-W_k^2+i\epsilon)},
\label{3h}
\end{equation}
where $W_k^2 \equiv |{\bf k}|^2 +m_\pi^2$.

Next we carry out the integration over $k_0$ by choosing the appropriate 
contour. From Eq.(\ref{3e}), when we calculate mass corrections from 
pion loops, we need the value of $\Sigma$ at $v \cdot p =m_0$. Since
$v \cdot p$ is a Lorentz scalar,
we are free to choose some special value of $v$ for this purpose.
With $v^0=1$ and ${\bf v}={\bf 0}$, and choosing the contour in the
lower half plane, in which there is only one pole for $k_0$, $W_k -i\epsilon$,
we arrive at the following expression:
\begin{equation}
X_1 =\frac{i}{6}\int \frac{{\rm d}^3 k}{(2\pi)^3} \frac{|{\bf k}|^2}
{W_k (W_k - \delta)}.
\label{3i}
\end{equation}
 
In Refs.\cite{lein1}-\cite{jones}, it has been argued that when the Compton
wavelength of the pion is smaller than the source of the pion, pion
loop contributions are suppressed as powers of $m_\pi / \Lambda$ where
$\Lambda$ characterizes the finite size of the source of the pion. We
follow this argument and introduce a cutoff $\Lambda$ in the integral
(\ref{3i}). Since the leading non-analytic contribution of pion loops
is associated with the infrared behavior of the integral, it does not
depend on the details of the cutoff. In the following, we will treat $\Lambda$ 
as a parameter to be fixed by lattice data.

With the cutoff $\Lambda$, we obtain the following result for
the integral (\ref{3i}):
\begin{eqnarray}
X_1 &=&\frac{i}{72\pi^2}\left\{12 (m_\pi^2-\delta^2)^{3/2}\left[
{\rm arctg}\frac{\Lambda+\sqrt{\Lambda^2+m_\pi^2}-\delta}{\sqrt{m_\pi^2
-\delta^2}}-{\rm arctg}\frac{m_\pi-\delta}{\sqrt{m_\pi^2
-\delta^2}}\right] \right. \nn\\
&&\left. +3\delta(2\delta^2-3 m_\pi^2) {\rm ln}
\frac{\Lambda+\sqrt{\Lambda^2+m_\pi^2}}{m_\pi}+3\delta\Lambda\sqrt{\Lambda^2
+m_\pi^2}
+6(\delta^2-m_\pi^2) \Lambda +2 \Lambda^3 \right\}, \nn\\
&&
\label{3j}
\end{eqnarray}
when $m_\pi^2 \geq \delta^2$;
\begin{eqnarray}
X_1 &=&\frac{i}{72\pi^2}\left\{6 (\delta^2 - m_\pi^2)^{3/2} {\rm ln}
\left(\frac{\Lambda+\sqrt{\Lambda^2+m_\pi^2}-\delta -\sqrt{\delta^2 - m_\pi^2}}
{\Lambda+\sqrt{\Lambda^2+m_\pi^2}-\delta +\sqrt{\delta^2 - m_\pi^2}}
\left|\frac{m_\pi -\delta +\sqrt{\delta^2 -m_\pi^2}}
{m_\pi -\delta -\sqrt{\delta^2 -m_\pi^2}}\right|\right) \right.\nn\\
&&\left. +3\delta(2\delta^2-3 m_\pi^2) {\rm ln}
\frac{\Lambda+\sqrt{\Lambda^2+m_\pi^2}}{m_\pi}+3\delta\Lambda\sqrt{\Lambda^2
+m_\pi^2}
+6(\delta^2-m_\pi^2) \Lambda +2 \Lambda^3 \right\},
\label{3k}
\end{eqnarray}
when $m_\pi^2 \leq \delta^2$.

In the case where $\delta=0$, we have
\begin{equation}
X_1 =\frac{i}{36\pi^2}\left(3m_\pi^3 {\rm arctg}\frac{\Lambda}{m_\pi}
-3 m_\pi^2 \Lambda +\Lambda^3 \right).
\label{3l}
\end{equation}

If we take the chiral limit $m_\pi \ra 0$ in Eq.(\ref{3k}), we can see that
the leading non-analytic term is
\begin{equation}
X_1|_{m_\pi \ra 0} =\frac{i}{32\pi^2}\frac{m_\pi^4}{\delta}{\rm ln}m.
\label{3k1}
\end{equation}
It can be easily checked that the same chiral limit behavior is obtained 
if we work with the dimensional regularization method in evaluating
$X^{\mu\nu}$. This is because the leading non-analytic contribution of the 
pion loops is only associated with their infrared behavior.

In Eq.(\ref{3e}), $m_0$ for a pseudoscalar heavy meson is $-\frac{3}{4}\Delta$,
and hence $\delta$ in Eq.(\ref{3f}) equals $-\Delta$. Therfore, for
pion loop contributions to the mass of a pseudoscalar heavy meson $P$, we have
\begin{eqnarray}
\sigma_P&=&-\frac{g^2}{16\pi^2 f_\pi^2}\left\{12 (m_\pi^2-\Delta^2)^{3/2}\left[
{\rm arctg}\frac{\Lambda+\sqrt{\Lambda^2+m_\pi^2}+\Delta}{\sqrt{m_\pi^2
-\Delta^2}}-{\rm arctg}\frac{m_\pi+\Delta}{\sqrt{m_\pi^2
-\Delta^2}}\right] \right. \nn\\
&&\left. -3\Delta(2\Delta^2-3 m_\pi^2) {\rm ln}
\frac{\Lambda+\sqrt{\Lambda^2+m_\pi^2}}{m_\pi}-3\Delta\Lambda\sqrt{\Lambda^2
+m_\pi^2}
+6(\Delta^2-m_\pi^2) \Lambda  \right. \nn\\
&& \left. +2 \Lambda^3 \right\},
\label{3m}
\end{eqnarray}
when $m_\pi^2 \geq \Delta^2$;
\begin{eqnarray}
\sigma_P&=&-\frac{g^2}{16\pi^2 f_\pi^2}\left\{ {\rm ln}
\left(\frac{\Lambda+\sqrt{\Lambda^2+m_\pi^2}+\Delta -\sqrt{\Delta^2 - m_\pi^2}}
{\Lambda+\sqrt{\Lambda^2+m_\pi^2}+\Delta +\sqrt{\Delta^2 - m_\pi^2}}
\left|\frac{m_\pi +\Delta +\sqrt{\Delta^2 -m_\pi^2}}
{m_\pi +\Delta -\sqrt{\Delta^2 -m_\pi^2}}\right|\right) \right.\nn\\
&&\left. \times 6 (\Delta^2 - m_\pi^2)^{3/2} -3\Delta(2\Delta^2-3 m_\pi^2) 
{\rm ln}
\frac{\Lambda+\sqrt{\Lambda^2+m_\pi^2}}{m_\pi}-3\Delta\Lambda\sqrt{\Lambda^2
+m_\pi^2} \right.\nn\\
&&\left. +6(\Delta^2-m_\pi^2) \Lambda 
+2 \Lambda^3 \right\},
\label{3n}
\end{eqnarray}
when $m_\pi^2 \leq \Delta^2$.

Now we turn to the pion loop corrections to heavy vector meson masses.
First we discuss Fig.2(b), where $P$ could be $D$ or $B$. This diagram
is caused by the $P P^* \pi$ vertices in Eq.(\ref{2t}).
In momentum space, Fig.2(b) can be expressed as
\begin{equation}
\frac{-i(g_{\mu\rho}-v_\mu v_\rho)}{2(v \cdot p -\frac{1}{4}\Delta)} 
(2i\Pi^{\rho\sigma})
\frac{-i(g_{\sigma\nu}-v_\sigma v_\nu)}{2(v \cdot p -\frac{1}{4}\Delta)},
\label{3o}
\end{equation}
where $p$ is the residual momentum of the heavy vector meson and 
$2i\Pi^{\rho\sigma}$ is given by the following integral:
\begin{equation}
2i\Pi^{\rho\sigma}=\frac{3 g^2}{f_\pi^2}\int \frac{{\rm d}^4 k}{(2\pi)^4}
\frac{k^\rho k^\sigma}{[v \cdot (p-k)+\frac{3}{4}\Delta](k^2-m_\pi^2)},
\label{3p}
\end{equation}
with $k$ the momentum of the pion in the loop. While we evaluate
Eq.(\ref{3p}), $\delta$ in Eq.(\ref{3f}) is equal to $v \cdot p 
+\frac{3}{4}\Delta$. Because of the factor $g_{\mu\rho}-v_\mu v_\rho$
in Eq.(\ref{3o}), only the $X_1$ term contributes. We define the 
coefficient of $g^{\rho\sigma}$ in $\Pi^{\rho\sigma}$ to be $\Pi$
\begin{equation}
\Pi^{\rho\sigma}|_{g^{\rho\sigma}}= \Pi,
\label{3q}
\end{equation}
so that after Fig.2(b) is included, the propagator of a heavy vector meson 
becomes proportional to Eq.({\ref{3c}), with $\Sigma$ being 
replaced by $\Pi$ and $m_0$ being
equal to $\frac{1}{4}\Delta$. While calculating the pion loop corrections to 
a heavy vector meson, we have to fix $v\cdot p =\frac{1}{4}\Delta$, as
required in Eq.(\ref{3e}). Consequently, $\delta$ in Eq.(\ref{3f}) is equal to 
$\Delta$. Then with the aid of Eqs.(\ref{3j}, \ref{3k}) we have
\begin{eqnarray}
\Pi &=&-\frac{g^2}{48\pi^2 f_\pi^2}\left\{12 (m_\pi^2-\Delta^2)^{3/2}\left[
{\rm arctg}\frac{\Lambda+\sqrt{\Lambda^2+m_\pi^2}-\Delta}{\sqrt{m_\pi^2
-\Delta^2}}-{\rm arctg}\frac{m_\pi-\Delta}{\sqrt{m_\pi^2
-\Delta^2}}\right] \right. \nn\\
&&\left. +3\Delta(2\Delta^2-3 m_\pi^2) {\rm ln}
\frac{\Lambda+\sqrt{\Lambda^2+m_\pi^2}}{m_\pi}+3\Delta\Lambda\sqrt{\Lambda^2
+m_\pi^2}
+6(\Delta^2-m_\pi^2) \Lambda +2 \Lambda^3 \right\},\nn\\
&&
\label{3r}
\end{eqnarray}
when $m_\pi^2 \geq \Delta^2$;
\begin{eqnarray}
\Pi &=&-\frac{g^2}{48\pi^2 f_\pi^2}\left\{ {\rm ln}
\left(\frac{\Lambda+\sqrt{\Lambda^2+m_\pi^2}-\Delta -\sqrt{\Delta^2 - m_\pi^2}}
{\Lambda+\sqrt{\Lambda^2+m_\pi^2}-\Delta +\sqrt{\Delta^2 - m_\pi^2}}
\left|\frac{m_\pi -\Delta +\sqrt{\Delta^2 -m_\pi^2}}
{m_\pi -\Delta -\sqrt{\Delta^2 -m_\pi^2}}\right|\right) \right.\nn\\
&&\left. \times 6 (\Delta^2 - m_\pi^2)^{3/2} +3\Delta(2\Delta^2-3 m_\pi^2) 
{\rm ln}
\frac{\Lambda+\sqrt{\Lambda^2+m_\pi^2}}{m_\pi}+3\Delta\Lambda\sqrt{\Lambda^2
+m_\pi^2} \right.\nn\\
&&\left. +6(\Delta^2-m_\pi^2) \Lambda 
+2 \Lambda^3 \right\},
\label{3s}
\end{eqnarray}
when $m_\pi^2 \leq \Delta^2$.

Now we discuss Fig.2(c), which arises from the $P^* P^* \pi$ vertices in 
the Lagrangian (\ref{2t}). In momentum space, Fig.2(c) can be expressed 
in the following explicit form:
\begin{eqnarray}
&&\frac{-i(g_{\mu\rho_1}-v_\mu v_{\rho_1})}{2(v \cdot p -\frac{1}{4}\Delta)} 
\left\{\frac{3g^2}{f_\pi^2}\epsilon^{\alpha_1 \nu_1 \rho_1 \sigma_1}
\epsilon^{\alpha_2 \nu_2 \rho_2 \sigma_2}(g_{\sigma_1 \rho_2}-v_{\sigma_1}
v_{\rho_2})v_{\nu_1}v_{\nu_2} \right. \nn\\
&&\left. \int \frac{{\rm d}^4 k}{(2\pi)^4}
\frac{k_{\alpha_1} k_{\alpha_2}}{[v \cdot (p-k)-\frac{1}{4}\Delta]
(k^2-m_\pi^2)} \right\} \frac{-i(g_{\nu\sigma_2}-v_\nu v_{\delta_2})}
{2(v \cdot p -\frac{1}{4}\Delta)}.
\label{3t}
\end{eqnarray}
After some algebra, this expression can be written in the form:
\begin{equation}
\frac{-i(g_{\mu\nu}-v_\mu v_{\nu})}{2(v \cdot p -\frac{1}{4}\Delta)}
\frac{i}{2(v \cdot p -\frac{1}{4}\Delta)}
\frac{6g^2}{f_\pi^2}X_1,
\label{3u}
\end{equation}
where in $X_1$, $\delta=v\cdot p -\frac{1}{4}\Delta$. 

Because of Fig.2(c), the propagator for a heavy vector meson becomes 
proportional to 
\begin{equation}
\frac{1}{v \cdot p -\frac{1}{4}\Delta - \Pi^{\prime}},
\label{3v}
\end{equation}
where 
\begin{equation}
\Pi^{\prime}=i\frac{3g^2}{f_\pi^2}X_1,
\label{3w}
\end{equation}
and again $\delta=v\cdot p -\frac{1}{4}\Delta$ in $X_1$. 
When we calculate corrections to a heavy vector meson mass
from Fig.2(c), $v \cdot p$ should be taken to be $\frac{1}{4}\Delta$,
as required by Eq.(\ref{3e}). Hence $\delta=0$. Using Eq.(\ref{3l})
we have 
\begin{equation}
\Pi^{\prime}=-\frac{g^2}{12 \pi^2 f_\pi^2}\left(3m_\pi^3 {\rm arctg}
\frac{\Lambda}{m_\pi}-3m_\pi^2\Lambda +\Lambda^3\right).
\label{3x}
\end{equation}

Adding $\Pi$ in Eqs.(\ref{3r}) and (\ref{3s}) 
and $\Pi^{\prime}$ in Eq.(\ref{3x})
together, we have the following expression for pion loop contributions
to the mass of a heavy vector meson $P^*$:
\begin{eqnarray}
\sigma_{P^*} &=&-\frac{g^2}{16\pi^2 f_\pi^2}
\left\{4 (m_\pi^2-\Delta^2)^{3/2}\left[
{\rm arctg}\frac{\Lambda+\sqrt{\Lambda^2+m_\pi^2}-\Delta}{\sqrt{m_\pi^2
-\Delta^2}}-{\rm arctg}\frac{m_\pi-\Delta}{\sqrt{m_\pi^2
-\Delta^2}}\right] \right. \nn\\
&&\left. +4m_\pi^3 {\rm arctg} \frac{\Lambda}{m_\pi}
+\Delta(2\Delta^2-3 m_\pi^2) {\rm ln}
\frac{\Lambda+\sqrt{\Lambda^2+m_\pi^2}}{m_\pi}+\Delta\Lambda\sqrt{\Lambda^2
+m_\pi^2}\right. \nn\\
&& \left. +2(\Delta^2-3m_\pi^2) \Lambda  
+2 \Lambda^3 \right\},
\label{3y}
\end{eqnarray}
when $m_\pi^2 \geq \Delta^2$;
\begin{eqnarray}
\sigma_{P^*} &=&-\frac{g^2}{16\pi^2 f_\pi^2}\left\{ {\rm ln}
\left(\frac{\Lambda+\sqrt{\Lambda^2+m_\pi^2}-\Delta -\sqrt{\Delta^2 - m_\pi^2}}
{\Lambda+\sqrt{\Lambda^2+m_\pi^2}-\Delta +\sqrt{\Delta^2 - m_\pi^2}}
\left|\frac{m_\pi -\Delta +\sqrt{\Delta^2 -m_\pi^2}}
{m_\pi -\Delta -\sqrt{\Delta^2 -m_\pi^2}}\right|\right) \right.\nn\\
&&\left. \times 2 (\Delta^2 - m_\pi^2)^{3/2} 
+4m_\pi^3 {\rm arctg} \frac{\Lambda}{m_\pi}+\Delta(2\Delta^2-3 m_\pi^2) 
{\rm ln}
\frac{\Lambda+\sqrt{\Lambda^2+m_\pi^2}}{m_\pi}\right.\nn\\
&&\left. +\Delta\Lambda\sqrt{\Lambda^2
+m_\pi^2} 
+2(\Delta^2-3m_\pi^2) \Lambda 
+2 \Lambda^3 \right\},
\label{3z}
\end{eqnarray}
when $m_\pi^2 \leq \Delta^2$.
 
Eqs.(\ref{3m}, \ref{3n}, \ref{3y}, \ref{3z}) are valid provided $m_\pi$ is not
far away from the chiral limit, i.e., when $m_\pi \le \Lambda$. It can be
seen from Eq.(\ref{3h}) that pion loop contributions vanish in the limit
$m_\pi \ra \infty$. In order to extrapolate lattice data from large pion mass
to the physical mass of the pion we need to know the behavior of heavy
meson masses at large $m_\pi$. As we know, a heavy meson is composed of 
a heavy quark and a light quark. In HQET, the heavy quark mass, $m_Q$,
is removed. Therefore, as the light quark mass, $m_q$, becomes large,
the heavy meson mass should be proportional to $m_q$. 
On the other hand, explicit lattice calculations show that over the range
of masses of interest to us, $m_\pi^2$ is proportional to $m_q$ \cite{lein1}.
Consequently, as $m_\pi$ becomes large, the heavy meson mass becomes
proportional to $m_\pi^2$. Based on these considerations, in order to 
extrapolate lattice data at large $m_\pi$ to the physical value of
the pion mass, we propose the following phenomenological functional 
form for the dependence of heavy meson masses on the mass of the pion
over the mass range of interest to us:
\begin{equation}
m_P=a_P + b_P m_\pi^2 +\sigma_P,
\label{3aa}
\end{equation}
for heavy pseudoscalar mesons, and
\begin{equation}
m_{P^*}=a_{P^*} + b_{P^*} m_\pi^2 +\sigma_{P^*},
\label{3bb}
\end{equation}
for heavy vector mesons. $\sigma_P$ and $\sigma_{P^*}$ are given in 
Eqs.(\ref{3m}, \ref{3n}) and Eqs.(\ref{3y}, \ref{3z}), respectively.
In next section, we will use this form to extrapolate lattice data
for heavy mesons.

Before turning to the lattice data we comment briefly on the explicit chiral 
symmetry breaking by the
terms in Eq.(\ref{2u}). Substituting Eqs.(\ref{2a}, \ref{2d}, \ref{2f},
\ref{2k}) into
Eq.(\ref{2u}) we have the following explicit expression for Eq.(\ref{2u}):
\begin{equation}
4\lambda_1 \sum_{a=1}^{3}m_{q^a}(P_{a \mu}^{* +}P_a^{* \mu} - P_a^+ P_a)
+4 \lambda_1^{\prime}\sum_{a=1}^{3}m_{q^a}
\sum_{a=1}^{3}(P_{a \mu}^{* +}P_a^{* \mu} - P_a^+ P_a),
\label{3cc}
\end{equation}
where we have made a Taylor expansion for $\xi$ and omitted $O(1/f_{\pi}^2)$
terms. It can be seen clearly that Eq.(\ref{3cc}) contributes equally to 
the mass of a heavy pseudoscalar meson
and that of a heavy vector meson. Therefore, it does not contribute to
their mass difference. Corrections to Eq.(\ref{3cc}) may modify the
propagators of heavy mesons to order $m_q O(1/f_{\pi}^2)$, with extra
suppression from $m_q$ with respect to the pion loop effects in
Eqs.(\ref{3m}, \ref{3n}) and Eqs.(\ref{3y}, \ref{3z}) and we will ignore
them.

\vspace{0.2in}
{\large\bf IV. Extrapolation of lattice data for heavy meson masses}
\vspace{0.2in}

The spectra of the $D$ and $B$ mesons were calculated in Ref.\cite{hein},
where NRQCD was used to treat the heavy quarks. In fact, when the mass
of a heavy quark, $m_Q$, is much larger than $\Lambda_{\rm QCD}$, it becomes
an irrelevant scale for the dynamics inside a heavy hadron. Consequently,
heavy meson states can be simulated on lattices with the aid of NRQCD
(where $m_Q$ is removed from the Hamiltonian) when the lattice spacing
is larger than the Compton wavelength of the heavy quark. Two values for 
the bare gauge coupling $\beta$, 5.7 and 6.2, were used. We choose to
use the data at $\beta=5.7$, since in this case the inverse lattice 
spacing $a^{-1}$ is about 1.116GeV, which is smaller than the masses of either
the $b$ or $c$ quarks. The box size is 2.1fm, corresponding to the volume
$12^3 \times 24$. In their simulations, three different values for the
hopping parameter $\kappa$, 0.1380, 0.1390, and 0.1400, were used.
The light quark mass is related to $\kappa$ as 
$m_q=\frac{1}{2a}(1/\kappa -1/\kappa_c)$, with $\kappa_c=0.1434$.

In our model there are three parameters to be fixed, $a_{P(P^*)}$, 
$b_{P(P^*)}$, and $\Lambda$ in Eqs.(\ref{3aa}) and 
(\ref{3bb}). These parameters are related to $g$ and $\lambda_2$
which represent interactions at the quark and gluon level and cannot be 
determined
from chiral perturbation theory for heavy mesons. As discussed in Section II,
they may depend slightly on $m_Q$. In our fitting procedure, we treat them as
effective parameters and assume that their $m_Q$ dependence has been
considered effectively. From the upper limit for the experimental decay width 
of $D^* \ra D \pi$ \cite{data} we have the upper limit $g^2 \le 0.5$
\cite{hlcpt1, wise2}. In our numerical work we let $g^2$ vary
from 0.3 to 0.5. From Eq.(\ref{2q}) $\lambda_2$ is related to the 
mass splitting
between a heavy vector meson and a heavy pseudoscalar meson. From experimental
data for the case of $B$ mesons, where $1/m_Q^2$ corrections can be
safely neglected, the value of $\lambda_2$ should be around -0.03GeV$^2$.
To see the dependence on $\lambda_2$ of our fits, we allow $\lambda_2$ 
to vary between -0.03GeV$^2$ and -0.02GeV$^2$.

Using the three simulation masses for $D$, $D^*$, $B$, and $B^*$ in
Ref.\cite{hein}, we fix the three parameters, $a_{P(P^*)}$, $b_{P(P^*)}$, 
and $\Lambda$ in Eqs.(\ref{3aa}) and (\ref{3bb})
with the least squares fitting method. In Tables 1 and 2, for different
values of $\lambda_2$ and $g^2$, we list the parameters $a_{P(P^*)}$, 
$b_{P(P^*)}$, and $\Lambda$ obtained by fitting the lattice data. Furthermore,
we give the masses of $D$, $D^*$, $B$, and $B^*$, and hyperfine 
splittings at the physical pion mass, $m_\pi^{\rm phys}$, after extrapolation.
With these parameters $a_{P(P^*)}$, $b_{P(P^*)}$, and $\Lambda$, we obtain
the masses of $D$, $D^*$, $B$, and $B^*$ as a function of the pion mass,
for different values of $\lambda_2$ and $g$. These results are shown in 
Figs.3 and 4. The mass differences between  $D$ ($B$) and $D^*$
($B^*$) as a function of the pion mass are shown in Figs.5 and 6. 
From these figures we can see that when the pion mass is smaller than about 
600MeV the extrapolations deviate significantly from linear behavior.
This is because the pion loop corrections begin to affect the
extrapolations around this point. As the pion mass becomes smaller
and smaller pion loop corrections become more and more important. 
Consequently, the dependence of these plots on $\lambda_2$ and $g^2$
is stronger when the pion mass is smaller. 

As discussed in Section III, the parameter $\Lambda$ characterizes the size
of the source of the pion. Since the sizes of $D$, $D^*$, $B$, and $B^*$
are different, we expect $\Lambda$ for them are different. This is
consistent with what is shown in Tables 1 and 2. Furthermore, we can see
that the $\Lambda$ difference between $D$ and $D^*$ is much bigger than
that between $B$ and $B^*$. This is because
the size difference between a heavy pseudoscalar meson and a heavy vector 
meson is caused by $1/m_Q$ effects. 

In the naive linear extrapolations pion loop corrections are ignored. Hence
the results do not depend on $\lambda_2$ and $g^2$. In Table 3 we list 
the results of the linear extrapolations for comparison.

It can be seen clearly that the mass differences between $D$ ($B$) and $D^*$
($B^*$) in our model 
are even smaller than those obtained in the naive linear extrapolations. Since
in the linear extrapolations, the hyperfine splittings at the physical mass of
the pion for $D$ and $B$ mesons are already smaller than experimental data,
the inclusion of pion loop effects makes the situation even worse. As
shown in Table 1, the hyperfine splitting between $D$ and $D^*$ is
$0.080 \sim 0.091$GeV in the range of our parameters, compared with the 
experimental value 0.14GeV, while
the result from the linear extrapolation is 0.114GeV in Table 3. 
Similarly, the hyperfine splitting between $B$ and $B^*$ in Table 2 is
$0.023 \sim 0.026$GeV in the range of our parameters, compared with 
the experimental value 0.046GeV and
the result from the linear extrapolation, 0.031GeV in Table 3.

In our fits, in addition to the uncertainties in the three parameters, 
$a_{P(P^*)}$, $b_{P(P^*)}$, and $\Lambda$, which are caused by 
the uncertainties of $\lambda_2$ and $g$, the errors in the lattice data can
also lead to some uncertainties in these three parameters. Since all the
three lattice data are at large pion masses, and since $\Lambda$ is mainly
related to the data at small pion masses, the error in our
determination of  $\Lambda$ is
very large. Besides $\Lambda$, $a_{P(P^*)}$ and  $b_{P(P^*)}$ also have some 
errors. In fact, the numerical results for $a_{P(P^*)}$, $b_{P(P^*)}$, 
and $\Lambda$
in Tables 1, 2, and 3 correspond to the central values of the lattice data.
As a consequence of heavy quark symmetry, the dynamics inside a
heavy meson does not depend on the mass of the heavy quark, so we expect that
the values of $\Lambda$ for $D$ and $B$ mesons are not very different from 
those of light mesons \cite{lein1}. 
Consequently, we expect that the values of 
$\Lambda$ in Tables 1 and 2 are quite reasonable. In our fits, we
find that the errors of the lattice data cause $1\% \sim 2\%$ relative
uncertainties for $a_{P(P^*)}$ ($P=D$ or $B$), $6\%$
for $b_D$, $9\%$ for $b_{D^*}$, $8\%$ for $b_B$, and $9\%$ for $b_{B^*}$.
These lead to about $1.3\%$ uncertainties for $m_D$, $1.5\%$ for $m_D^*$,
$1.2\%$ for $m_B$, and $1.2\%$ for $m_{B^*}$. As a result, the relative
uncertainties of the hyperfine splittings are about $13\%$ and $54\%$ in the 
cases of $D$ and $B$ mesons, respectively. In spite of all these errors,
our conclusion that the hyperfine splitting obtained after a careful
treatment of chiral corrections are
smaller than those obtained using naive linear extrapolation, remains
unchanged.

\begin{table}
\caption{Fitted parameters, masses of $D$ and $D^*$ and their difference
at $m_\pi^{\rm phys}$ (Numbers without (with) brackets are for $D$ ($D^*$))}
\begin{center}
\begin{tabular}{lcccc}
\hline
\hline
 &$\lambda_2$=-0.02GeV$^2$ &$\lambda_2$=-0.02GeV$^2$  
&$\lambda_2$=-0.03GeV$^2$ &$\lambda_2$=-0.03GeV$^2$  \\ 
 &$g^2=0.3$ &$g^2=0.5$ &$g^2=0.3$ &$g^2=0.5$ \\
\hline
$\Lambda ({\rm GeV})$ &0.43 [0.65] &0.38 [0.55] &0.45 [0.63] &0.39 [0.55]\\
\hline
$a_{D[D^*]} ({\rm GeV})$ &0.5439 [0.6848] &0.5436 [0.6796] &0.5444 [0.6821]
&0.5438 [0.6807]\\
\hline
$b_{D[D^*]} ({\rm GeV}^{-1})$ &0.2082 [0.1783] &0.2083 [0.1809] 
&0.2078 [0.1798] 
&0.2082 [0.1800] \\
\hline
$m_{D[D^*]}^{\rm fit} ({\rm GeV})$ &0.5374 [0.6282] &0.5364 [0.6233] 
&0.5375 [0.6254] 
&0.5369 [0.6166]\\
\hline
$m_{D^*}^{\rm fit}-m_{D}^{\rm fit} ({\rm GeV})$ &0.0908 &0.0869 &0.0879 
&0.0796\\
\hline
\hline
\end{tabular}
\end{center}
\end{table}

\begin{table}
\caption{Fitted parameters, masses of $B$ and $B^*$ and their difference
at $m_\pi^{\rm phys}$ (Numbers without (with) brackets are for $B$ ($B^*$))}
\begin{center}
\begin{tabular}{lcccc}
\hline
\hline
 &$\lambda_2$=-0.02GeV$^2$ &$\lambda_2$=-0.02GeV$^2$  
&$\lambda_2$=-0.03GeV$^2$ &$\lambda_2$=-0.03GeV$^2$  \\ 
 &$g^2=0.3$ &$g^2=0.5$ &$g^2=0.3$ &$g^2=0.5$ \\
\hline
$\Lambda ({\rm GeV})$ &0.62 [0.65] &0.53 [0.56] &0.62 [0.65] &0.54 [0.56]\\
\hline
$a_{B[B^*]} ({\rm GeV})$ &0.7540 [0.7915] &0.7507 [0.7884] &0.7535 [0.7918]
&0.7519 [0.7887]\\
\hline
$b_{B[B^*]} ({\rm GeV}^{-1})$ &0.1657 [0.1581] &0.1673 [0.1593] 
&0.1661 [0.1579] 
&0.1665 [0.1591] \\
\hline
$m_{B[B^*]}^{\rm fit} ({\rm GeV})$ &0.7138 [0.7394] &0.7107 [0.7342] 
&0.7146 [0.7389] 
&0.7107 [0.7336]\\
\hline
$m_{B^*}^{\rm fit}-m_{B}^{\rm fit} ({\rm GeV})$ &0.0256 &0.0236 &0.0242 
&0.0229\\
\hline
\hline
\end{tabular}
\end{center}
\end{table}

\begin{table}
\caption{Fitted parameters, masses of $P$ and $P^*$ and their difference
at $m_\pi^{\rm phys}$ for linear extrapolation (Numbers without (with) 
brackets are for $P$ ($P^*$))}
\begin{center}
\begin{tabular}{cccc}
\hline
\hline
$a_{D[D^*]} ({\rm GeV})$ & $b_{D[D^*]} ({\rm GeV}^{-1})$ 
& $m_{D[D^*]}^{\rm fit} ({\rm GeV})$ 
&$m_{D^*}^{\rm fit}-m_{D}^{\rm fit} ({\rm GeV})$ \\
\hline
0.5397 [0.6540] &0.2112 [0.1995] & 0.5438 [0.6579] & 0.1141 \\
\hline
\hline
$a_{B[B^*]} ({\rm GeV})$ & $b_{B[B^*]} ({\rm GeV}^{-1})$ 
& $m_{B[B^*]}^{\rm fit} ({\rm GeV})$ 
&$m_{B^*}^{\rm fit}-m_{B}^{\rm fit} ({\rm GeV})$ \\
\hline
0.7311 [0.7621] & 0.1812 [0.1780] & 0.7346 [0.7656]& 0.0310\\
\hline
\hline
\end{tabular}
\end{center}
\end{table}

\vspace{0.6in}
{\large\bf V. Summary and discussion}
\vspace{0.2in}

QCD possesses chiral symmetry when light quark masses go to zero and heavy 
quark symmetry when heavy quark masses go to infinity. Combining these
two symmetries leads to chiral perturbation theory for heavy mesons
which are invariant under both chiral symmetry and heavy quark symmetry.
We have evaluated pion loop corrections to heavy meson propagators
with the aid of chiral perturbation theory for heavy mesons as the 
Compton wavelength of the pion becomes larger than the size of the 
heavy mesons.
This leads to the dependence of heavy meson masses on the pion mass.
In order to study hyperfine splittings, we took the color-magnetic-moment 
operator at order $1/m_Q$ in HQET 
into account. This operator breaks heavy quark
spin symmetry and is primarily responsible for the mass difference between a 
heavy pseudoscalar meson and a heavy vector meson. The small masses of the 
light quarks break chiral symmetry explicitly. We showed that 
contributions to the mass difference between a heavy pseudoscalar meson and a 
heavy vector meson from these terms are suppressed by light quark masses
with respect to the pion loop contributions we considered in 
chiral perturbation 
theory. When $m_\pi$ becomes large, lattice data show that heavy
meson masses are proportional to $m_\pi^2$. Based on these considerations, 
we proposed a phenomenological functional form with three parameters to 
extrapolate the lattice data. Because it guarantees the model independent
chiral behavior of QCD, our model is more appropriate 
than a naive linear extrapolation.
The parameters in our model are fixed by the least 
squares fitting method, while fitting the lattice data for the masses of 
heavy mesons $D$, $D^*$, $B$, and $B^*$ in the large pion mass region 
($m_{\pi} \ge 680$MeV). It is found that the 
hyperfine splittings extrapolated in this way are even smaller 
than those obtained in the linear extrapolations, in which 
the extrapolated hyperfine splittings for both $D$ and $B$ mesons are 
already smaller than the experimental data. 

There are some uncertainties in our model. We have two parameters,
$\lambda_2$ and $g$, which are related to the color-magnetic-moment 
operator at order $1/m_Q$ in HQET and the interactions between heavy
mesons and Goldstone bosons in chiral perturbation theory, respectively.
In the ranges we choose for these two parameters, we have about $13\%$
and $11\%$ uncertainties for the hyperfine splitting in the $D$ and $B$ cases,
respectively. Furthermore, the errors in the lattice data also lead to some
uncertainties when we fix the three parameters $a_{P(P^*)}$, $b_{P(P^*)}$, 
and $\Lambda$ in our model. Since all the lattice data are at high
pion masses, the error in $\Lambda$ is very large. However, we believe that
the range of $\Lambda$ we obtained is appropriate because of 
considerations based on 
heavy quark symmetry. The errors for $a_{P(P^*)}$ and $b_{P(P^*)}$
lead to about $13\%$ and $54\%$ uncertainties for hyperfine splittings 
in the cases of $D$ and $B$ mesons, respectively. Despite all these
uncertainties, the hyperfine splittings obtained in our model are
smaller than those in the naive linear extrapolations. Our analysis
shows that the current lattice data for hyperfine splittings at large pion 
masses are probably too small to give hyperfine splittings at
the physical pion mass which are consistent with experiments. 

Some approximations made in current lattice simulations may be the cause 
of these small 
hyperfine splittings. The quenched approximation might be one reason.
In fact, in lattice simulations for light spectroscopy the hyperfine
splittings are also too small \cite{yoshie}. Furthermore, as pointed out in 
Ref.\cite{hein}, the lattice results for hyperfine splittings are sensitive
to the coefficient of the ${\bf \sigma\cdot B}$ term in NRQCD 
which is at order $1/m_Q$ and which is the leading term to cause 
hyperfine splittings. The inclusion of radiative corrections beyond
tadpole improvement for this coefficient may increase hyperfine splittings.
Another reason might be the light quark mass dependence of the clover 
coefficient in the clover action for light quarks. In addition, finite
size effects and higher order terms in NRQCD in lattice simulations
may lead to an underestimate of hyperfine splittings at large light quark 
mass as well. More careful lattice simulations with more data and better
accuracy are urgently needed to resolve this important problem.

\vspace{2cm}

\noindent {\bf Acknowledgment}:

This work was supported by the Australian Research Council.

\newpage

\vspace{0.2in}

\noindent{\large \bf Figure Captions} \\
\vspace{0.4cm}

\noindent Fig.1 Pion loop corrections to the propagator of heavy 
pseudoscalar meson $P$ ($P$ could be $D$ or $B$).
\vspace{0.2cm}

\noindent Fig.2 Pion loop corrections to the propagator of heavy 
vector meson $P^*$ ($P$ could be $D$ or $B$).
\vspace{0.2cm}

\noindent Fig.3 Phenomenological fits to the lattice data for the masses
of $D^*$ (the upper lines) and $D$ (the lower lines) as a function of the 
pion mass.  
The solid (dashed) lines correspond to $\lambda_2=-0.02$GeV$^2$ and 
$g^2=0.3$ ($g^2=0.5$). The dot (dot dashed) lines correspond to 
$\lambda_2=-0.03$GeV$^2$ and $g^2=0.3$ ($g^2=0.5$). 
\vspace{0.2cm}

\noindent Fig.4 Phenomenological fits to the lattice data for the masses
of $B^*$ (the upper lines) and $B$ (the lower lines) as a function of the 
pion mass. 
The solid (dashed) lines correspond to $\lambda_2=-0.02$GeV$^2$ and 
$g^2=0.3$ ($g^2=0.5$). The dot (dot dashed) lines correspond to 
$\lambda_2=-0.03$GeV$^2$ and $g^2=0.3$ ($g^2=0.5$). 
\vspace{0.2cm}

\noindent Fig.5 Phenomenological fits to the lattice data for the hyperfine
splitting between $D^*$ and $D$ as a function of the pion mass. 
The solid (dashed) lines correspond to $\lambda_2=-0.02$GeV$^2$ and 
$g^2=0.3$ ($g^2=0.5$). The dot (dot dashed) lines correspond to 
$\lambda_2=-0.03$GeV$^2$ and $g^2=0.3$ ($g^2=0.5$). 
\vspace{0.2cm}

\noindent Fig.6 Phenomenological fits to the lattice data for the hyperfine
splitting between $B^*$ and $B$ as a function of the pion mass. 
The solid (dashed) lines correspond to $\lambda_2=-0.02$GeV$^2$ and 
$g^2=0.3$ ($g^2=0.5$). The dot (dot dashed) lines correspond to 
$\lambda_2=-0.03$GeV$^2$ and $g^2=0.3$ ($g^2=0.5$). 
\vspace{0.2cm}

\newpage

\begin{figure}[p]
\begin{center}
{\epsfsize=13.7in\epsfbox{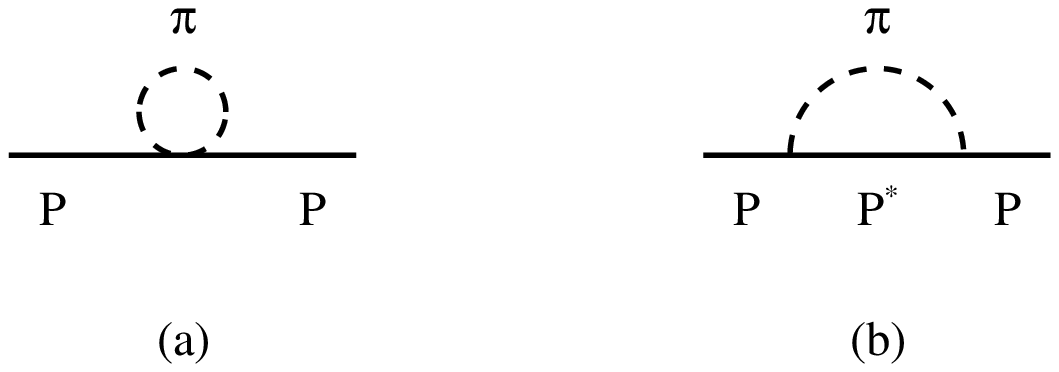}}
\end{center}
\vspace{1.5cm}
\centerline{Fig.1}
\end{figure}
\vspace{2cm}

\begin{figure}[p]
\begin{center}
{\epsfsize=15in\epsfbox{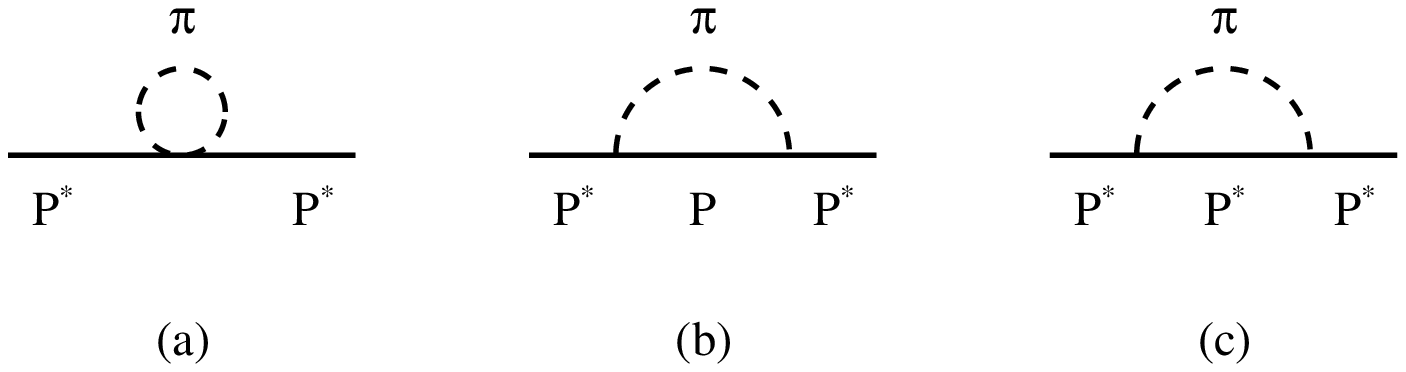}}
\end{center}
\vspace{1.5cm}
\centerline{Fig.2}
\end{figure}
\vspace{2cm}

\newpage

\begin{figure}[p]
\begin{center}
{\epsfsize=14.5in\epsfbox{lacpv.eps}}
\end{center}
\vspace{0.cm}
\centerline{Fig.3}
\end{figure}

\begin{figure}[p]
\begin{center}
{\epsfsize=14.5in\epsfbox{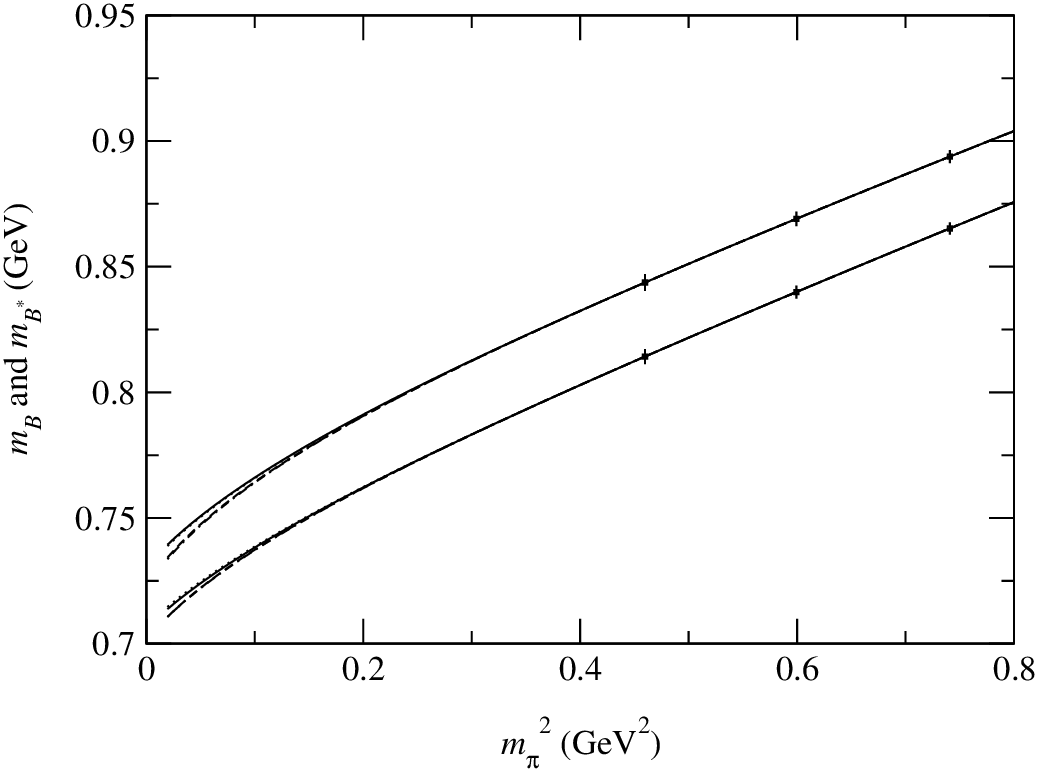}}
\end{center}
\vspace{0.cm}
\centerline{Fig.4}
\end{figure}

\newpage

\begin{figure}[p]
\begin{center}
{\epsfsize=14.5in\epsfbox{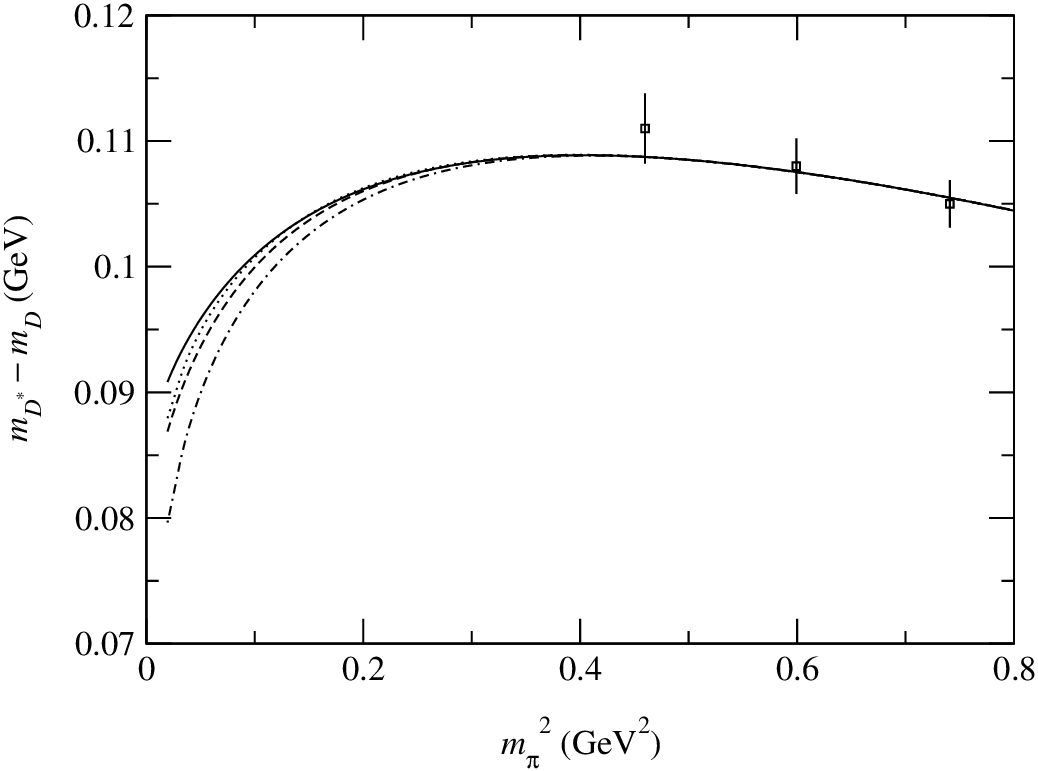}}
\end{center}
\vspace{0.cm}
\centerline{Fig.5}
\end{figure}

\begin{figure}[p]
\begin{center}
{\epsfsize=14.5in\epsfbox{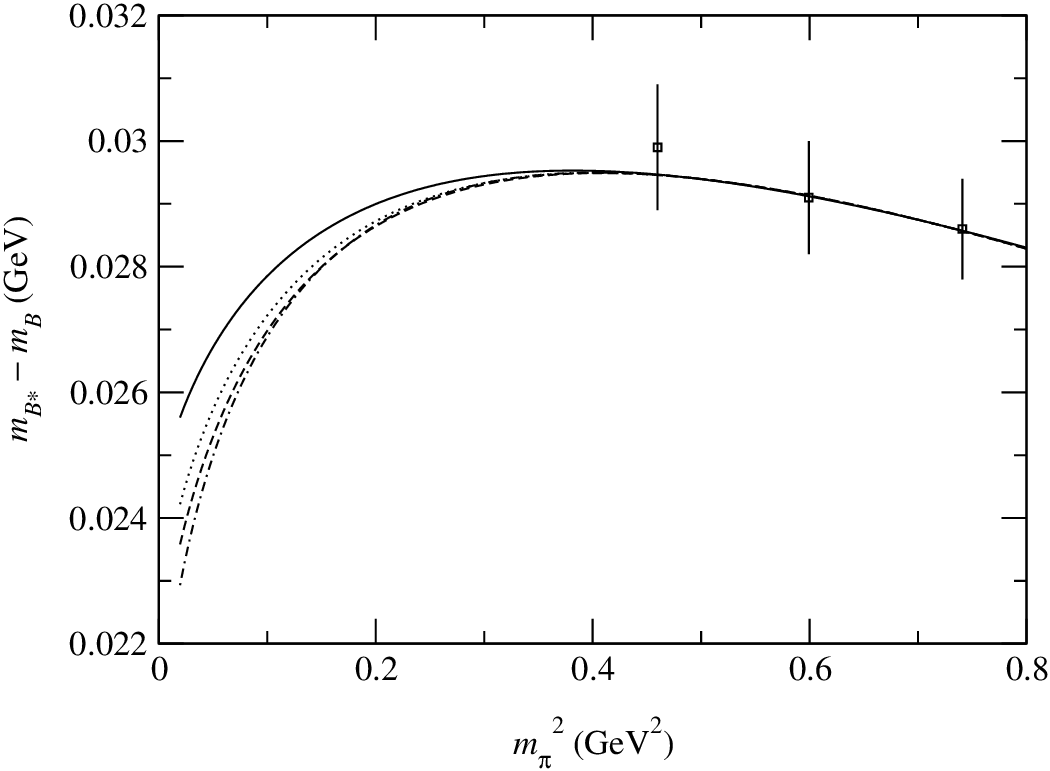}}
\end{center}
\vspace{0.cm}
\centerline{Fig.6}
\end{figure}


\newpage


\baselineskip=20pt

\begin{thebibliography}{7}

\bibitem{light} T. Bhattacharya, R. Gupta, G. Kilcup, and S. Sharpe, 
Phys. Rev. {\bf D53} (1996) 6486; CP-PACS Collaboration, S. Aoki {\it et
al.}, Phys. Rev. {\bf D60} (1999) 114508; UKQCD Collaboration, C. Allton 
{\it et  al.}, Phys. Rev. {\bf D60} (1999) 034507; UKQCD Collaboration, 
K. Bowler {\it et  al.}, Phys. Rev. {\bf D62} (2000) 054506.

\bibitem{khan1} A. Ali Khan {\it et  al.}, Phys. Rev. {\bf D62} (2000) 054505.

\bibitem{hein} J. Hein {\it et  al.}, Phys. Rev. {\bf D62} (2000) 074503.

\bibitem{khan2} A. Ali Khan {\it et  al.}, Phys. Rev. {\bf D56} (1997) 7012.

\bibitem{gockeler} M. G\"{o}ckeler {\it et al.}, Phys. Rev. {\bf D53} 
(1996) 2317; M. G\"{o}ckeler {\it et al.}, Nucl. Phys. Proc.
Suppl. {\bf 53} (1997) 81; D. Dolgov {\it et al.}, Nucl. Phys. Proc.
Suppl. {\bf 94} (2001) 303.

\bibitem{lacagnina} UKQCD Collaboration, G. Lacagnina, hep-lat/0109006.

\bibitem{wise}  N. Isgur and M.B. Wise, Phys. Lett.  {\bf B232} (1989) 113,
{\bf B237} (1990) 527; H. Georgi, Phys. Lett. {\bf B264} (1991) 447;
see also M. Neubert, Phys. Rep. {\bf 245} (1994) 259 for the review.

\bibitem{hlcpt1} M.B. Wise, Phys. Rev. {\bf D45} (1992) 2188. 

\bibitem{hlcpt2} G. Burdman and
J. Donoghue, Phys. Lett.  {\bf B280} (1992) 287; T. Yan, H. Cheng,
C. Cheung, G. Lin, and H. Yu, Phys. Rev. {\bf D46} (1992) 1148;
B. Grinstein, E. Jenkins, A. Manohar, M. Savage, and M. Wise, Nucl.
Phys. {\bf B380} (1992) 369; A. Falk, Phys. Lett.  {\bf B305} (1993) 268;
A. Falk and B. Grinstein, Nucl. Phys. {\bf B416} (1994) 771;
P. Ko, Phys. Rev. {\bf D47} (1993) 1964; P. Cho, Nucl. Phys. {\bf B396} (1993)
183; E. Jenkins, M. Luke, A. Manohar, and M. Savage, Nucl. Phys. {\bf B397} 
(1993) 84; Y.-B. Dai, X.-H. Guo, C.-S. Huang, and H.-Y. Jin,
Commun. Theor. Phys. {\bf 24} (1995) 453.

\bibitem{lein1} D.B. Leinweber, A.W. Thomas, K. Tsushima, and S.V. Wright,
Phys. Rev. {\bf D61} (2000) 074502; Phys. Rev. {\bf D64} (2001) 094502.

\bibitem{lein2} D.B. Leinweber, D.H. Lu, and A.W. Thomas, Phys. Rev. 
{\bf D60} (1999) 034014; E.J. Hackett-Jones, D.B. Leinweber, and A.W. Thomas,
Phys. Lett.  {\bf B489} (2000) 143; D.B. Leinweber and A.W. Thomas,
Phys. Rev. {\bf D62} (2000) 074505.

\bibitem{detmold} W. Detmold, W. Melnitchouk, and A.W. Thomas, 
Eur. Phys. J. {\bf C13} (2001) 1;
W. Detmold, W. Melnitchouk, J.W. Negele, D.B. Renner, and 
A.W. Thomas, Phys. Rev. Lett. {\bf 87} (2001) 172001. 

\bibitem{jones} E.J. Hackett-Jones, D.B. Leinweber, and A.W. Thomas,
Phys. Lett.  {\bf B494} (2000) 89.

\bibitem{data} The Particle Data Group, D.E. Groom {\it et al.},
Eur. Phys. J. {\bf C15} (2000) 1.

\bibitem{wise2} M.B. Wise, hep-ph/9306277,
Lectures given at CCAST Symposium on Particle Physics at the Fermi Scale, 
1993. Published in CCAST Symposium (1993), 71. 

\bibitem{yoshie} T. Yoshie, Nucl. Phys. B (Proc. Suppl.) {\bf 63} (1998) 3.
\end{thebibliography}
\end{document}